\documentclass[a4paper,11pt]{article}
\pdfoutput=1 % if your are submitting a pdflatex (i.e. if you have
\usepackage{jheppub} % for details on the use of the package, please
\usepackage[T1]{fontenc} % if needed
\usepackage{graphics}
\usepackage{amsfonts}
\usepackage{amsmath}
\usepackage{graphicx}
\usepackage{epic,eepic,epsfig}
\usepackage{dcolumn}
\usepackage{bm}
\usepackage{color}
\usepackage{bbm}
\newcommand{\nn}{\nonumber}
\newcommand{\bea}{\begin{eqnarray}}
	\newcommand{\ena}{\end{eqnarray}}

\DeclareMathOperator{\id}{id} 
 \DeclareMathOperator{\tr}{tr}

\title{\boldmath Quantum nonequilibrium dynamics from Knizhnik-Zamolodchikov equations}

\author[a]{Tigran A. Sedrakyan,}
\author[b]{Hrachya M. Babujian}

	\affiliation[a]{Department of Physics, University of Massachusetts, Amherst, Massachusetts 01003, USA}
	\affiliation[b]{ A. Alikhanyan National Scientific Laboratory, Yerevan Physics Institute, Yerevan,  Armenia}

\emailAdd{tsedrakyan@umass.edu}
\emailAdd{babujian@yerphi.am}

\abstract{We consider a set of non-stationary quantum models. We show that their dynamics can be studied using links to Knizhnik-Zamolodchikov (KZ) equations for correlation functions in conformal field theories. We specifically consider the boundary  Wess-Zumino-Novikov-Witten model, where equations for correlators of primary fields are defined by an extension of KZ equations and explore the links to dynamical systems. As an example of the workability of the proposed method, we provide an exact solution to a dynamical system that is a specific multi-level generalization of the two-level Landau-Zenner system known in the literature as the Demkov-Osherov model. The method can be used to study the nonequilibrium dynamics in various multi-level systems from the solution of the corresponding KZ equations.}

\begin{document} 
\maketitle
\flushbottom

\section{\bf Introduction} 

There has been considerable interest in nonequilibrium dynamics of quantum systems (see e.g., Refs.~\cite{Kamen,smith})
%describing the transition dynamics with with time-dependent Hamiltonian operators that varie so that the energy difference between the two states is a linear function of time
%,~\cite{Sierra_RMP, Poghossyan-2004,teodorescu,Demkov-1967,Demkov-2000,bow-tie-1997,Demkov-2001,Delft1, Yuzbash,garcia,caux,  Yuzbashyan-2015, Yuzbashyan-2017, Sedrakyan-2010, Altland-2007,Altland-2008, Sinitsin-2018,Sinitsin-2021} 
over the last couple of decades that surged again recently in connection with quantum information problems \cite{QI0,QI1,TV,QI7,QI6,Scip2,QI8,QI3,QI11,QI12,QI14,QI15,QI4,QI5,QI9,QI10,QI2}.
These include the dynamical discrete-state Bardeen-Cooper-Shriffer (BCS) pairing models  \cite{Rich1,Rich2,Delft1,Poghossyan-2004,Sierra_RMP,Yuzbash1,Yuzbash2,teodorescu,garcia,caux,Sedrakyan-2010,grits,TV,Scip4,Scip3}, where for example the interaction strength can be made time dependent,  and various multi-level Landau-Zenner tunneling models and their many body generalizations \cite{Demkov-1967,bow-tie-1997,Demkov-2000,Demkov-2001,Sinitsyn1,Altland-2007,Altland-2008,Yuzbashyan-2015,Yuzbashyan-2017,Raikh,Sinitsin-2018,Scip1,Sinitsin-2021}. 

The cornerstone in the integrable properties of the discrete-state BCS hamiltonian is Richardson's exact solution\cite{Rich1,Rich2} and its extensions to a class of models with far-reaching applications in the theory of strongly correlated quantum matter. The exact solution is achieved by employing the connections to an exactly solvable Gaudin magnet \cite{Gaudin-1976}. The connection is established upon rewriting the bilinears of spinful fermion operators in $s=1/2$ operators using Schwinger fermion construction. As a result, the Richardson Hamiltonian is mapped to an exactly solvable spin-1/2 magnet with mutually commuting integrals of motion. Notably, the number of the integrals of motion is equal to the number of degrees of freedom in the system, a property that guarantees integrability. Moreover, these integrals of motion of the Richardson model are a linear combination of the corresponding integral of motion of the Gaudin magnet and $S_l^3$ component of the corresponding $s=1/2$ operator.

Another important connection is between the Gaudin magnet and a two-dimensional conformal theory\cite{CFT} (CFT) associated with the $SU(2)$ group, namely the Wess-Zumino-Novikov-Witten (WZNW) theory. Generally, it is well established that theories, where current algebra is a fundamental symmetry, are governed by WZNW action, which has topological properties. The connection with magnet is through the correlation functions of the primary fields in CFT that satisfy a series of Knizhnik-Zamolodchikov (KZ)\cite{KZ} equations involving constants of motion of the Gaudin model. The correlation functions in $SU(2)$ WZNW have been obtained from the solution of KZ equations with constants of motion of the Gaudin model by one of us and a collaborator in Refs.~\cite{YerPr,Babujian-1994,Babujian-1996}. This approach was further developed in Ref.~\cite{Sedrakyan-2010}, where a boundary $SU(2)$ WZNW model was identified and it was shown that the correlation functions of the primary fields there satisfy the series of modified KZ equations, where the Gaudin operators are replaced by constants of motion of the Richardson model. The approach was also used to obtain exact solutions to a variety of dynamical systems in Ref.~\cite{Sedrakyan-2010}. 
Notably, in Ref.~\cite{Vgho} the method of Refs.~\cite{YerPr,Babujian-1994,Babujian-1996} was applied to a broad class of Gaudin magnets with spin operators, $S_n$, being the generators of most simple complex Lie algebras \footnote{The statement applies to all simple complex Lie algebras except those of $E_8$, $G_2$, and $F_4.$}. 
A different method for diagonalization of Gaudin Hamiltonians was developed in Ref.~\cite{Feigin}. Here one employs a bosonization approach (using the bosonic representation of the corresponding affine algebra) and treats the diagonalization problem and KZ equations on the same footing. 
As revealed in Refs.~\cite{Feigin,F3}, certain ordinary differential operators called opers could encode the spectrum of the quantum Gaudin Hamiltonians corresponding to a simple Lie algebra. Amazingly, the opers that carry all the information about the spectra of the Gaudin model are in direct correspondence with the Langlands dual/twisted Lie algebra (rather than the simple Lie algebra itself). This can be regarded as a particular case of the geometric Langlands correspondence \cite{F1,BD,FK}

Other profound implications of KZ equations in gauge theories and quantum spin chains have recently been reported in Refs.~\cite{nekrasov1,nekrasov2}.  Bethe states and KZ equations were thoroughly investigated for generalized $sl(2)$ Gaudin models in \cite{Salom19, Salom}. Moreover,  a relation between integrable Kondo problems in products of chiral $SU(2)$ WZNW models and affine $SU(2) $Gaudin models was recently reported in \cite{Gaiotto}. Another relationship between the quantum Gaudin models with boundary and the classical many-body integrable systems of Calogero-Moser type was established in \cite{Zabrodin}. More interesting correspondences between Gaudin models, conformal field theories, and BCS -Richardson pairing models are reported in Refs.~\cite{Manojlovic,Skrypnyk,Delduc,Lacroix2,Lacroix1,Lukyanov}.

{In the last 30 years, interesting, purely mathematical approaches have been developed to construct integrable models linked to loop groups and current algebras. 
In Refs.~\cite{Adams-1988, Adams-1990, Harnad-1994},  a momentum map, namely a Hamiltonian (Lie) group action on a symplectic manifold, was used to construct conserved quantities for the action in the context of KZ equations. It gave a dual formulation of integrable Hamiltonian systems as isospectral flows in the two-loop algebras. The momentum map of Poisson manifold of rank-$k$ perturbations of a fixed $n \times n$ matrices to dual of pairs of loop algebras $\tilde{gl}(k)$ and $\tilde{gl}(n)$ presented a base for generating integrable Gaudin type of models linked to current algebra.
	
In Refs.~\cite{Tarasov-2000, Tarasov-2004},  a holonomic system of differential equations with rational coefficients was found, which are compatible with an extension of the rational KZ equations by an element of the Cartan subalgebra. These equations were termed rational dynamical differential (DD) equations. Note that the first such extension of KZ equations by $\sigma^z$ in $SU(2)$ case was formulated in Ref.~\cite{Babujian-Kitaev} for the Maxwell-Blokh system. 
Further, in Ref.~ \cite{Tarasov-2004} it was shown,
that $(gl_k, gl_n)$ duality, which plays an important role in the representation theory and the classical invariant theory,\cite{Zhelobenko-1983, Howe-1995}, gives rise a
duality between KZ equations (for the Lie algebra $gl_n$) and DD equations for $gl_k$.

A comprehensive study of dualities between various quantum and classical many-body systems, including dualities of KZ equations,
 with application to the calculation of integrable probabilities in the stochastic process are presented in Ref.~\cite{Gorsky-2021}.
 The probabilities in the stochastic process are treated within the conventional machinery of integrable models, including transfer matrices and Bethe ansatz equations.
 An interesting KZ interpretation of duality in stochastic processes is discussed in \cite{Chen-2020}.
}

Therefore, strikingly it turns out that the pairing Hamiltonian and a boundary WZNW CFT are related to one another in a nontrivial manner via the modification of KZ equations for primary fields in CFT that include integrals of motion of the pairing Hamiltonian. Consequently, it is expected that analytic properties of the correlation functions of conformal field theories and the spectral and dynamical properties of the pairing Hamiltonian can be explored on the same footing. One of the aims of the present work is to manifest this relation explicitly. To this end, we discuss the precise modification of the KZ equations and that these are satisfied by the correlation functions of primary fields in the boundary WZNW model. The main goal of the present work, however, is to study several dynamical systems that can be solved by employing the connection through modified KZ equations, where time-dependence is explicit, but can be treated exactly upon solving the set of KZ equations.

As we see, through nontrivial correspondences, seemingly disconnected domains of theoretical physics are linked one with another. Following the same prescription above, several dynamical systems discussed below can be mapped on a variant of the pairing Hamiltonian, with the interaction parameter playing the role of a function of the time variable. Then, it can be shown that the Schr\"{o}dinger equations for wave functions in these models lead to the specific extension of KZ equations for correlation functions of primary fields in SU(2) current algebra. Using the relation to the KZ equations, one can find the solutions of such dynamical models leading to various incarnations of the modified KZ equations in different situations. 
To this end, in the second part of the paper, we consider first the multi-level generalization of the two-level Landau-Zenner system, namely the Demkov-Osherov model, and its alternations. We show that the  Demkov-Osherov model and its partner generalized bow-tie models are linked to an extension of KZ equations, which can be solved exactly. We present this analytical solution and discuss its implications.

The remainder of the paper is organized as follows. In section II, we construct the boundary WZNW model, which leads to an extended version of KZ equations.
In Section III, we consider the dynamical multi-level Landau-Zenner problem and its simplest realizations: the Demkov-Osherov model and its modification. A class of exact solutions to these dynamical systems is presented. In the Appendix, we present the details of the construction of the boundary term in the WZNW model.

\section{KZ equations  for  Boundary WZNW model}
Another connection is between the gapless generalization of the Heisenberg model to spin-$s$  \cite{Babujian-1982, Babujian-1984, Takhtajian-1982}. Specifically, it was shown by Affleck and Haldane,\cite{Affleck-1986} that the connection reveals itself in identifying the quantum field theory corresponding to the critical point with the WZNW model with action \cite{PW,PW2}
\bea
\label{WZNW}
S_{WZNW}(g)=\frac{\mathsf{k}}{16\pi}\int\limits_{S^2} dz d\bar{z}  
\tr\left[\partial_{a}g^{\dagger}\partial^{a}g \right] \qquad \qquad \\
-\frac{i \mathsf{k}}{24 \pi}\int\limits_{\Sigma^3} d^3 x \varepsilon ^{\mu\nu\rho}\tr
\left[ g^{\dagger}\partial _{\mu}g g^{\dagger}\partial _{\nu}g g^{\dagger}\partial _{\rho}g \right].\quad\nn
\ena
Here integration in the first term is over a two-dimensional manifold, $S^2$, corresponding to a compactified complex plane parameterized by $(w,\bar{w})$ and group elements $g(w,\bar{w}) \in SU(2)$. Integration in the second topological WZNW term is over a three-dimensional manifold $\Sigma^3$, with $x = (w,\bar{w},\xi) \in \Sigma^3$, whose boundary at $\xi = 0$ is the aforementioned sphere, $S^2 = \partial \Sigma^3$ and the function $g(w,\bar{w};\xi=0)$ is extended into the interior of the ball $\xi \in \left[0,\,1\right]$ in a non-unique way. The parameter $\mathsf{k}$ in Eq.~(\ref{WZNW}) is an integer number, which is the level of the corresponding CFT and linked to spin $s$ of the chain model as $\mathsf{k}=2s$. The WZNW action is invariant under conformal and non-Abelian current algebras. 

In the seminal work Ref.~\cite{KZ}, Knizhnik and Zamolodchikov have shown that
 $N$-point correlation functions, $G(w_1\ldots w_N\mid \bar{w}_1\ldots \bar{w}_N ) = \left\langle \phi^{\mu_1}_{s_1}(w_1,\bar{w}_1) \ldots \phi^{\mu_N}_{s_N}(w_N,\bar{w}_N) \right\rangle_{S_{\rm WZNW}}$ of primary fields  with spins $0 \leq s_i \leq \mathsf{k}/2,\; i=1, \cdots N $ of the WZNW model Eq.~(\ref{WZNW}), satisfy a system of first-order differential
KZ equations. In the holomorphic sector these equations can be written as
\bea
\label{GKZ}
\left[ (\mathsf{k}+2)\partial_{w_l} - \hat{ H}_l^G \right] G\left(\{w_l\}\right) = 0, 
\ena
where
\bea
\label{Gauden}
 \hat{H}_l^G =\sum\limits_{l \ne l'}\frac{\hat{{\bf S}}_l \cdot \hat{{\bf S}}_{l'}}{w_l-w_{l'}}
\ena
are the integrals of motion of the Gaudin magnet model \cite{Gaudin-1976} with $\hat{{\bf S}}_l $ being a set of generators of $SU(2)$ with $l=\overline{1N}$.
Gaudin model Hamiltonian $\hat{H}^G$ is integrable and is linear combination of integrals of motion $ \hat{H}^G=2 \sum_{l=1}^N w_l \hat{H}_l^G$.
As we pointed in the introduction, a class of models including the celebrated Richardon's pairing Hamiltonian\cite{Rich1,Rich2,Delft1,Poghossyan-2004,Sierra_RMP,Yuzbash1,Yuzbash2,teodorescu,garcia,caux,Sedrakyan-2010,grits,TV,Scip4,Scip3}, whose conserved integrals of motion can be regarded as a generalization of those of the Gaudin magnet\cite{Amico,amico2,italians}
\bea
\label{MKZ-H}
\hat{H}_l^R=\lambda S_l^3+ \hat{H}_l^G,\quad \big[\hat{H}_l^R,\hat{H}_{l'}^R\big]=0,
\ena
leading to {extended by $\lambda S_l^3$ term} KZ equations (EKZ). Namely,
\bea
\label{MKZ}
\left[ (\mathsf{k}+2)\partial_{w_l} - \hat{ H}_l^R \right] G\left(\{w_l\}\right) = 0.
\ena 
Below we will follow the idea put forward in Ref.~\cite{Sedrakyan-2010} and show that EKZ equations (\ref{MKZ}), with $l=1\ldots N$, are satisfied by the primary fields of the boundary WZNW CFT. 

\subsection{Boundary WZNW term}

\begin{figure}[t]
	\centerline{\includegraphics[width=60mm,angle=0,clip]{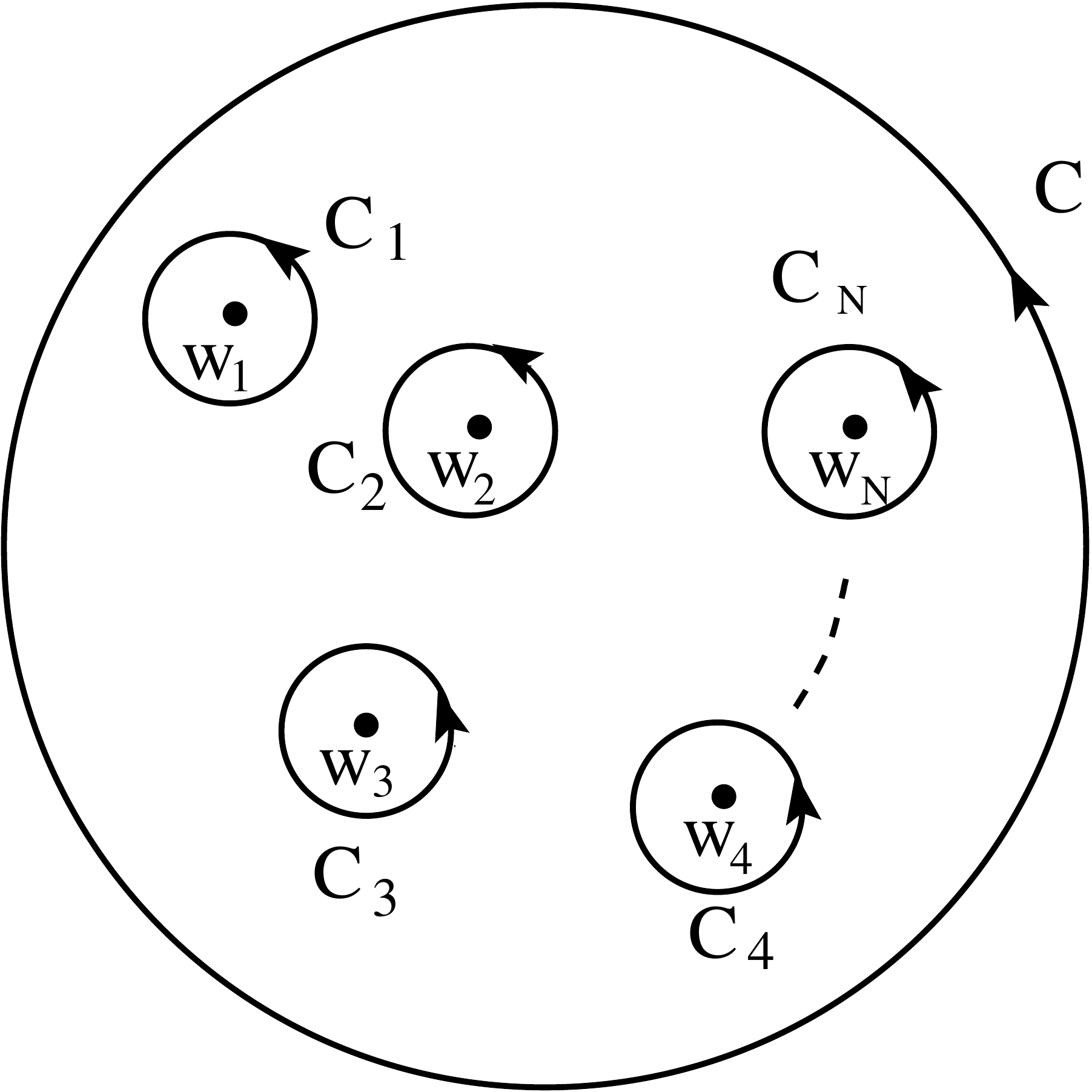}}
	\caption{Boundary contour {\cal C} corresponds to the limit of large distances. It shrinks to the sum of contours $C_i, \; i=1,\cdots N$ encircling points $w_i$: ${\cal C}=\bigsqcup_{j=1}^N C_i$.    } 
	\label{contours}
\end{figure}

Consider left and right boundary terms of the WZNW model based on left-flowing  and right-flowing  currents $J^a(w)=\tr[\hat{\bf S}^a g^\dagger(w)\partial_w g(w)], \; \bar{J}^a(\bar{w})=\tr[\hat{\bf S}^a g^\dagger(\bar{w})\partial_{\bar{w}} g(\bar{w})]\; a=1,2,3$, with spin-$s$ generators of $SU(2)$ algebra, $\hat{\bf S}^a$, and
\bea
\label{B1}
S^L_{bound}(\cal C)&=& \alpha \oint_{\cal C} 
d w w {J}^3(w), \nn\\
S^R_{bound}(\bar{\cal C})&=& \alpha \oint_{\bar {\cal C}} d\bar{w} \bar{w} {\bar J}^3(\bar{w}).
\ena
Here the coefficient $\alpha\in{\mathcal R}$ is real.
Then, upon adding these currents to the action of the WZNW model, $S_{WZNW}(g)$, we obtain the model dubbed boundary  WZNW (BWZNW):
\bea
\label{MWZNW}
S_{BWZNW}(g)=S_{WZNW}(g)+S^L_{bound}({\cal C})+
S^R_{bound}(\bar{\cal C}).
\ena 
We show below that the primary fields in this boundary theory satisfy  EKZ equations Eq.~(\ref{MKZ}). In Eq.~(\ref{B1}) contours $\cal C$ and $\bar{\cal C}$
are largest contours encompassing complex numbers $w_l$ and 
approaching ${\cal C}_\infty$. Contour ${\bar {\cal C}}$ have clockwise
rotation. Both contours should contain all points $w_l$ of the correlators, which
are under consideration (see Fig.\ref{contours}).
Note also that due to conformal invariance, the left currents $J^a(w)$, $a=1,2,3$, 
do not depend on ${\bar w}$ and likewise the right currents, ${\bar J}^a({\bar w})$, do not depend on $w$. 

Generally, we are interested in calculating a correlation function of arbitrary   spin $0\leq s_i\leq (\mathsf{k}/2)$ primary fields~\cite{CFT} at points
$w_i, i=1\ldots N$ in the $SU(2)$ boundary WZNW model (see Fig.\ref{contours}). Namely $G(w_1,\cdots w_N)=\langle \phi_{s_1}(w_1)\ldots \phi_{s_N}(z_N\rangle _{S_{\rm BWZNW}}$, where  functional averaging is defined in a standard way 
\begin{eqnarray}
\langle \phi_{s_1}(w_1)\ldots \phi_{s_N}(w_N\rangle _S=\int \left[{\cal D}\phi\right]  \phi_{s_1}(w_1)\cdots \phi_{s_N}(w_N) e^{-S_{\rm WZNW}-S_{bound}({\cal C})}. 
\end{eqnarray}
Due to this relation,
one can evaluate any correlator in the boundary model via its relation to the a correlation function of the same primary fields multiplied with $\Phi\left[{\cal C}\right]=e^{-S_{bound}({\cal C})}$ but with average in the bulk WZNW model. Namely,
\begin{eqnarray} 
G(z_1,\cdots z_N)=
\langle \Phi\left[{\cal{C}}\right] \phi_{s_1}(z_1)\cdots \phi_{s_N}(z_N) \rangle _{S_{\rm WZNW}}.
\end{eqnarray}
%{\color{red} Here, $\Phi\left[{\cal{C}}\right]$ can be taken out of the expectation value and applied to the correlation function of the WZNW model $\langle \phi_{s_1}(z_1)\cdots \phi_{s_N}(z_N) \rangle$ from the outside. 

%%%%%%%%%%%%%%
%%%%%%%%%%%%%% 
With this rearrangement and conceptually new averaging procedure, one can show that the primary fields satisfy the EKZ equation following the standard procedure of deriving the KZ equations in the ordinary WZNW model.  We present a detailed discussion of the outlined procedure in the next subsection. 

\subsection{KZ equations in the presence of the boundary term} 
The emergence of ordinary KZ equations in CFT is the consequence of the fact that by definition, primary fields $\phi_{s}(w)$ are null vectors in current algebra $SU(2)$. 
Namely the primary fields 
fulfill constraints $\big[L_0-\frac{1}{\mathsf{k}+2} J_{-1}^a J_0^a\big]\phi_{s}(w)=0$,
where operators $J_{0}^a$ and $J_{-1}^a$ 
%are acting on the first primary field in the correlation function $\langle %\phi_{s_1} \cdots \phi_{s_N}\rangle$ 
are $0$ and $-1$ coefficients of the Laurent series of the current $J^a(w)$, while $L_0 =\partial_w$ is the zero component of the Virasoro algebra. When they are acting on any field at $w\in C_1$, then one has 
\bea
\label{JJ}
J_{0}^a=\oint_{C_1}J^a(u) du,\quad 
J_{-1}^a=\oint_{C_1} \frac{J^a(u)}{u-w} du.
\ena
The null vector condition is the same also in the boundary model. Therefore its gives the following equation for correlation functions
\bea
\label{null vector}
&&\langle \big[ \partial_{w_1}-\frac{1}{\mathsf{k}+2}J^a_{-1}J_0^a\big] \phi_{s_1}(w_1)
\cdots \phi_{s_N}(w_N) \rangle_{\text{\tiny{BWZNW}}}=\nn \\ 
&& \langle e^{\Phi(\cal C)}\big[ \partial_{w_1}-\frac{1}{\mathsf{k}+2}J^a_{-1}J_0^a\big] \phi_{s_1}(w_1)
\cdots \phi_{s_N}(w_N) \rangle_{\text{\tiny{WZNW}}}\nn \\ 
&&=0.
\ena
Here we consider equation for the first primary field with
argument $w_1$, therefore operator $J^a_{-1}J_0^a$ is acting on
$\phi_{s_1}(w_1)$. This means that in the integral representation
of these operators, Eq.~(\ref{JJ}), contours should circle $w_1$.
As one can see, in Eq.~(\ref{null vector}), there are two terms: The first term is
the derivative, $\partial_{w_1}$. The second term is $J_{-1}^a J_0^a$. By definition, the primary fields are those that are eigenstates of the current operator,
$J^a(u)$. Namely, they fulfill the following operator algebra relation
\bea
\label{PF}
J^a(u) \phi_{s_i}(w_i)= \frac{S_i^a}{u-w_i} \phi_{s_i}(w_i)+ \cdots.
\ena 

Upon using this relation, the direct calculation of two terms in the null vector condition (\ref{null vector}) for the first primary field at $w_1$ gives 
(see details in Appendix A)
\bea
\label{term-1}
&&\langle e^{\alpha \oint_{\cal C} d w w {J}^3(w)} \partial_{w_1} \phi_{s_1}(w_1)
\cdots \phi_{s_N}(w_N)\rangle_{\text{\tiny{WZNW}}}\nn\\
&=&\partial_{w_1} \langle  e^{\alpha \sum_{i=1}^N w_i S_i^3} \phi_{s_1}(w_1)
\cdots \phi_{s_N}(w_N)\rangle_{\text{\tiny{WZNW}}}
\ena
%\bea
%\label{term-1}
%&&\langle e^{\alpha \oint_{\cal C} d w w {J}^3(w)} \partial_{w_1} \phi_{s_1}(w_1)
%\cdots \phi_{s_N}(w_N)\rangle_{\text{\tiny{WZNW}}}\nn\\
%&=&e^{\alpha w_1 S_1^3}\partial_{w_1}\langle \phi_{s_1}(w_1)
%\cdots \phi_{s_N}(w_N)\rangle_{\text{\tiny{WZNW}}}
%\ena
and
\bea
\label{term-2}
&&\langle e^{\alpha \oint_{\cal C} d w w {J}^3(w)} J_{-1}^a J_0^a \phi_{s_1}(w_1)
\cdots \phi_{s_N}(w_N)\rangle_{\text{\tiny{WZNW}}}\nn\\
&=&\sum_{j=2}^N \frac{S_j^a S_1^a}{w_1-w_j}e^{\alpha \sum_{i=2}^N w_i S_i^3}\langle \phi_{s_1}(w_1)
\cdots \phi_{s_N}(w_N)\rangle_{\text{\tiny{WZNW}}}\\
&+&c_1 \alpha S_1^3 e^{\alpha \sum_{i=2}^N w_i S_i^3}\langle \phi_{s_1}(w_1)
\cdots \phi_{s_N}(w_N)\rangle_{\text{\tiny{WZNW}}},\nn
\ena
where $c_1=S_1^a S_1^a$ and the summation over repeating indices, $a=1,2,3$, is assumed (see Appendix A for all pertinent calculations of the present subsection).
Combining these two equations into the null-vector condition, Eq.~(\ref{null vector}),  for the correlation
function 
\bea
\label{corF}
G(w_1,\cdots w_N)=\langle \phi_{s_1}\cdots \phi_{s_n}\rangle_{\text{\tiny{BWZNW}}}\nn\\
=\langle e^{\alpha \sum_{i=2}^N w_i S_i^3} \phi_{s_1}(w_1)
\cdots \phi_{s_N}(w_N)\rangle_{\text{\tiny{WZNW}}},
\ena
where we have used operator algebra (\ref{PF}) for primary field, we arrive
at following EKZ equations
\bea
\label{MKZ-2}
\Big[(\mathsf{k}+2)\partial_{w_1}-\lambda S_1^3-\sum_{j=2}^N \frac{S_j^aS_1^a}{w_1-w_j}\Big]
G(w_1,\cdots w_N)=0.\nn\\
\ena
Here the coefficient $\lambda =\alpha c_1$.  Such an equation was established in Ref.~\cite{Babujian-Kitaev} for the Maxwell-Bloch system. Below we consider a class of nonequilibrium systems whose dynamics can be studied from the exact solution of the corresponding EKZ equations.  

\section{Multi-level Landau-Zenner problem and its descendants}

In a related context, certain integrable aspects of models with time-dependent Hamiltonians have recently been studied in Refs.~\cite{Sedrakyan-2010,Yuzbashyan-2017,Sinitsin-2018,Sinitsin-2021}.
Here we present the solutions of the Demkov-Osherov\cite{Demkov-1967}, bow-tie (BT) \cite{bow-tie-1997} and generalized version of the bow-tie (GBT) models\cite{Demkov-2000, Demkov-2001}. This is achieved from the exact solutions of the associated EKZ equations. We show that GBT contains a set of integrals of motion, which form EKZ for correlation functions, where $n-2$ spins should be considered classical or, equivalently, large-spins $s\gg 1$. Using the off-shell Algebraic Bethe Ansatz (OSABA) technique developed in Refs.~\cite{Babujian-1994, Babujian-1996, Babujian-Kitaev}, we find the solution of this EKZ for correlation functions in GBT.

In its general formulation, the multilevel LZ problem is defined by a time-dependent Hamiltonian of the form\bea
\label{multi-LZ}
H_{LZ}=\hat{A}+\hat{B} t,
\ena
where $\hat{A}$ and $\hat{B}$ are $(n+1) \times (n+1)$ Hermitian matrices. 
In general, for arbitrary $\hat{A}$ and $\hat{B}$ matrices, 
corresponding time-dependent Schr\"{o}dinger equation, 
\bea
\label{SE}
-i \partial_t \Psi(t)=H_{LZ}\Psi(t),
\ena   
can be solved formally via utilizing the $T$-exponent representation of the wave function. The latter however  is practically not very useful for computation of observables and other applications. Though, in some simple cases, a solution can be found.  The most simple case is when matrix $\hat{B}$ has $n$ coinciding eigenvalues, $b_2$, and only one is different, $b_1\neq b_2$. Then the unitary matrix $U$, which diagonalizes $\hat{B}$, is degenerate on a subgroup of $n \times n$ unitary matrices. The latter can be used to diagonalize $n \times n$ minor of the matrix $\hat{A}$.  To simplify the described $t$-dependent Hamiltonian, one can
subtract from the Hamiltonian Eq.~(\ref{LZ1}) the diagonal matrix, $b_2t\times \id $, and rescale the time by the factor $b_1-b_2$.
After this procedure, the Hamiltonian $H_{LZ}$ acquires the form 
\bea
\label{LZ1}
H_{DO}= \left( 
\begin{array}{ccccccc}
	t+a_{00}& v_{01}&\cdot&\cdot&\cdot&v_{0n}\\
	v_{01}& a_{01}& 0 & \cdot & \cdot & 0\\
	v_{02}&0& a_{02}&0 &\cdot& 0\\
	\cdot&\cdot&\cdot&\cdot &\cdot & \cdot \\
	\cdot&\cdot&\cdot&\cdot &\cdot & \cdot \\
	v_{0 n-1} &0&\cdot&\cdot & a_{0 n-1}& 0\\
	v_{0n} & 0 &\cdot  &\cdot  & 0 & a_{0 n}
\end{array}
\right),
\ena 
which was first defined in Ref.~\cite{Demkov-1967} and subsequently dubbed Demkov-Osherov (DO)
model.
BT model (parametrized by a set of constants $\{r_i\}$) is equivalent to the DO model and their Hamiltonians
are linked by the following linear transformation:
\bea
\label{BT} 
a_{0i}=r_i t +t.
\ena
%where $H_{DO}$ is the Hamiltonian of Demkov-Osherov model.
These models and their generalizations were discussed in Refs.~\cite{Yuzbashyan-2015,Yuzbashyan-2017,Raikh}.

\subsection{The solution of Demkov-Osherov and bow-tie models}

We look for the eigenvalues and wave functions of the Hamiltonian, $H_{DO} \mid x_m(t)\rangle = E_m^0(t) \mid x_m(t)\rangle $, in the form
\bea
\label{EV}
\mid x_m(t)\rangle=\sum_{k=0}^n \frac{\gamma_k}{x_m(t)-\epsilon_k} \mid k\rangle,
\ena
where
\bea
\label{k}
\mid k \rangle =(0,\cdots 1, \cdots 0)^T,\;\text{and}\quad k=0,1,\cdots n.
\ena
Here in ket $\mid k \rangle$ the unity entry stands at the $k$'th place where the leftmost entry has number zero and we have introduced a set of yet unknown parameters, $(\gamma_0; \{\gamma_i\})$, $i=1,\ldots, N$. Now, we assume that the energy eigenvalues corresponding to wave function, $\mid x_m(t)\rangle$, are of the form
	\bea
	\label{energy}
	E^0_m(t)=\frac{\gamma_0^2}{x_m(t)-\epsilon_0},
	\ena
where the functions $x_m(t)$ and the connection between $\gamma_k$ and $\epsilon_k$ and parameters entering the DO Hamiltonian Eq.~(\ref{LZ1}) are yet to be determined from self-consistency relations. This so far is an ansatz, which will simplify the final result considerably. In this basis, the DO Hamiltonian (\ref{LZ1}) acquires a simple form
\bea
\label{DO}
H_{DO}&=&(t+a_{00})|0\rangle \langle 0 |+\sum_{i=1}^n a_{0i} |i\rangle \langle i |\nn\\ 
&+& \sum_{i=1}^n v_{0i} \big(|0\rangle \langle i |+|i\rangle \langle 0 | \big).
\ena   
We substitute now the expressions in Eq.~(\ref{EV}) for eigenstate and Eq.~(\ref{energy})
for energy into the Schr\"{o}dinger equation. This yields
\bea
\label{SchrEq}
H_{DO} \mid x_m(t)\rangle = E_m^0(t) \mid x_m(t)\rangle.
\ena
Consider the equation corresponding to basis element $| k=0 \rangle$.
The equation thus will acquire the following algebraic form:
\bea
\label{EV2}
&&(a_{00}+t)\frac{\gamma_0}{x_m(t)-\epsilon_0}+\sum_i \frac{v_{0i}\gamma_i}{x_m(t)-\epsilon_i}\nn\\
&=&\frac{\gamma_0^2}{x_m(t)-\epsilon_0}\cdot \frac{\gamma_0}{x_m(t)-\epsilon_0}.
\ena
For basis elements $| i \rangle$, with $i=1,\ldots,n$, the Schr\"{o}dinger equation Eq.~(\ref{SchrEq}) yields
\bea
\label{EV22}
\frac{a_{0i} \gamma_i}{x_m(t)-\epsilon_i} + \frac{v_{0i} \gamma_0}{x_m(t)-\epsilon_0}
= \frac{ \gamma_0^2}{x_m(t)-\epsilon_0} \frac{\gamma_i}{x_m(t)-\epsilon_i}.
\ena

It is straightforward to check that the eigenvalue equations Eqs.~(\ref{EV2})
and (\ref{EV22}) will be
fulfilled if the original parameters $v_{01}, \cdots v_{0n}$ and $a_{01}, \cdots a_{0n}$
are self-consistently connected with $\gamma_i$  and $\epsilon_i-\epsilon_0$ for $i=1,\ldots, n$ as
\bea
\label{connections}
v_{0i}&=&\frac{\gamma_0 \gamma_i}{\epsilon_0-\epsilon_i},\nn\\
a_{0i}&=&\frac{\gamma_0^2}{\epsilon_i-\epsilon_0}, \qquad i=1,\cdots n.
\ena
From Eq.~(\ref{EV2}), the parameters $a_{00}$ and the time variable, $t$, are bound with relations
\bea
\label{a00}
a_{00}=\sum_{i=1}^n \frac{\gamma_i^2}{\epsilon_i-\epsilon_0}, \\
\label{time}
t=\sum_{i=0}^n \frac{\gamma_i^2}{x_m(t)-\epsilon_i}.
\ena
Eq.~(\ref{a00}) yields a condition under which the eigenvalue equation has a solution, while Eq.~(\ref{time}) for time gives polynomial equation of $n+1$ order for $x_m(t)$. That polynomial $t$-dependent equation defines $m=0,\cdots n$ solutions in the form of functions for $x_m(t)$.  Importantly, the relation Eq.~(\ref{time}) appears when we equate the time dependent terms in the left and right hand sides of Eq. (\ref{EV2}).

% Equations for $| k \rangle$ with  $k\neq 0$ are also fulfilled due to relations in Eq.~(\ref{connections}).
From Eqs.~(\ref{connections}) and (\ref{a00}), one can find
\bea
\label{av}
\frac{v_{0i}}{a_{0i}}& =&- \frac{\gamma_i}{\gamma_0} , \;\; i=1,\cdots, n,\\
\label{av1}
a_{00}&=&\sum_{i}^{n}\frac{v_{0i}^2}{a_{0i}}.
\ena
These equations directly relate parameters in the DO Hamiltonian Eq.~(\ref{LZ1}), with the set of new parameters $(\gamma_0; \{\gamma_i\})$. 
Note that the relation (\ref{av1}) does not essentially restrict the DO Hamiltonian (\ref{LZ1}). Upon starting from the Hamiltonian with arbitrary diagonal entries, $a^{\prime}_{0i},$,  $i=0,1,\cdots, n$, and making a constant shift $a_{0i}=a^\prime_{0i}+a$, one will always fulfill the condition
Eq.~(\ref{av1}) by solving the polynomial equation of $n+1$ order for $a$. This procedure is equivalent to shifting the Hamiltonian by a constant, which does not change anything. 
So for the general class of polynomials that support real solutions for $a$, Eq.~(\ref{av1}) does not limit the general DO model and its solution.

Relations (\ref{connections}-\ref{time}) show that we can always absorb the parameter $\epsilon_0$ into $\epsilon_i$, $i=1,\cdots, n$ and $x_m(t)$ as $\epsilon_i -\epsilon_0 \rightarrow \epsilon_i $ and $x_m(t)-\epsilon_0 \rightarrow x_m(t)$. Nevertheless, we keep it for completeness. After these remarks, it is clear that transformation of parameters $(a_{0i}, v_{0i}, t) \rightarrow (\epsilon_i, \gamma_i, \gamma_0)$ is essentially exact and retains the correct number of degrees of freedom.

\subsection{Demkov-Osherov model and its alternations: The link to KZ equations}

DO, and BT models can be altered by redefining the properties of eigenvalues of the matrix $\hat{B}$. 
Let us assume that the matrix $\hat{B}$
has $n-1$ identical eigenvalues, $b_2$, and two other eigenvalues, $b_1$. 
Then the diagonalizing unitary matrix, $U$, will have degeneracy over subgroup of $SU(n-1)$ matrices, which will allow diagonalizing $(n-1) \times (n-1)$ diagonal minor of $\hat{A}$.
Then, subtracting from the Hamiltonian Eq.~(\ref{multi-LZ})  the diagonal matrix, $b_2t\times \id $, and after appropriate rescaling of the time by the factor $b_1-b_2$,
%\bea
%\label{LZj}
%H_1= \left( 
%\begin{array}{cccccccc}
%	a_{10}& v_{10}&0&\cdot&0&\cdot&\cdot&0\\
%	v_{10} &t+a_{11} & v_{11}& \cdot & v_{1j}& \cdot& \cdot & v_{1n}\\
%	0& v_{11}& \cdot & \cdot & \cdot & \cdot &\cdot&\cdot\\
%	\cdot&\cdot&\cdot&\cdot &\cdot & \cdot& \cdot& \cdot \\
%	0&v_{1j}& \cdot&\cdot &t+a_{1j}& \cdot&\cdot&0\\
%	\cdot&\cdot&\cdot&\cdot &\cdot & \cdot& \cdot&\cdot \\
%	\cdot &\cdot&\cdot&\cdot &\cdot&\cdot& a_{1\; n-1}& 0\\
%	0 & v_{1n} &\cdot  &\cdot  & 0 &\cdot&0& a_{1n}
%\end{array}
%\right),
%\ena 
one arrives at the  following alternation of Demkov-Osherov model (ADO) given in the form of a simple matrix Hamiltonian:
\bea
\label{LZj}
H_{\text{\tiny ADO}}= \left( 
\begin{array}{cccccccc}
	t+v_{00}& v_{01}&v_{02}&\cdot&v_{0j}&\cdot&\cdot&v_{0n}\\
	v_{01} &t+v_{11} & v_{12}& \cdot & v_{1j}& \cdot& \cdot & v_{1n}\\
	v_{02}& v_{12}& a_{2} & 0 & 0 & \cdot &\cdot&0\\
	\cdot&\cdot&0&\cdot &\cdot & \cdot& \cdot& \cdot \\
	v_{0j}&v_{1j}& 0&\cdot &a_{j}& \cdot&\cdot&0\\
	\cdot&\cdot&\cdot&\cdot &\cdot & \cdot& \cdot&\cdot \\
	\cdot &\cdot&\cdot&\cdot &\cdot&\cdot& a_{ n-1}& 0\\
	v_{0n}& v_{1n} &0 &\cdot  & 0 &\cdot&0& a_{n}
\end{array}
\right)\;\;
\ena 
Since the Hamiltonian is Hermitian, all parameters, $v_{ij}$  and $a_j$, $j=2, \cdots n$, that are present in it, are real. In the bra-ket notations, the Hamiltonian Eq.~(\ref{LZj})
reads
\bea
\label{DO-2}
&&\hspace{-1cm}H_{\text{\tiny ADO}}=\sum_{k=0,1}(t+v_{kk})|k\rangle \langle k |+\sum_{i=2}^n a_{i} |i\rangle \langle i |\nn\\ 
&&\hspace{-1cm}+v_{01}(|0\rangle \langle 1 |+|1\rangle \langle 0 |)+ \sum_{k=0,1;i=2}^n v_{ki} \big(|k\rangle \langle i |+|i\rangle \langle k | \big).
\ena   
In Ref.~\cite{Raikh}, the model was studied semiclassically for $n=3$, within the independent level crossing approximation. 

Below we are going to link the time-dependent Schr\"{o}dinger equation $-i \partial_t \Psi=H_{\text{\tiny GBT}}\Psi$, with $\Psi=(\psi_0,\psi_1,\cdots \psi_n)^T$, with EKZ equations and show that the solution of the latter satisfies the former. To this end, after Fourier transformation, the Schr\"{o}dinger equation
acquires the form of $\omega \Psi= H_{\omega}\Psi$. The latter in detailed form reads
\bea
\label{Sch-2}
\omega \psi_0&=&(-i \partial_\omega+v_{00})\psi_0+v_{01}\psi_1 + \sum_{k=2}^n v_{0k}\psi_k\nn\\
\omega \psi_1&=&(-i \partial_\omega+v_{11})\psi_1+v_{01}\psi_0 + \sum_{k=2}^n v_{1k}\psi_k\\
\omega \psi_k&=&a_k \psi_k + v_{0k} \psi_0 + v_{1k}\psi_1,\quad k=2,\cdots n. \nn
\ena
The solution of the last set of equations for $\psi_k$ with $k=2, \cdots n$ is straightforward, and yields
\bea
\label{set}
\psi_k=\frac{v_{0k}\psi_0+v_{1k}\psi_1}{\omega-a_k}, \quad k=2,\cdots n.
\ena
Substituting these expressions for $\psi_k$ into the first two equations in 
(\ref{Sch-2}), we obtain two equations for $\psi_{0,1}$. The latter  can be simplified further by the following transformation of the wave functions $\psi_{0,1}=e^{i \omega^2/2} \bar{\psi}_{0,1}$. After this transformation, the linear in $\omega$ term can be eliminated from the equations. Finally, we see that these two equations reduce to the following $2 \times 2$ matrix equation
\bea
\label{2x2}
i\partial_{\omega}\Phi=\sum_{\mu=0}^3\Big[b_1^\mu S^\mu_1+\sum_{k=2}^n\frac{ S_1^\mu b_k^\mu}{\omega-a_k}\Big]\Phi,\quad \Phi=(\bar\psi_0,\bar\psi_1)^T
\ena  
where the matrix $S_1^\mu=(\text{1},\sigma_1,\sigma_2,\sigma_3),\;\mu=0,1,2,3$ is given by the unity and the set of three Pauli matrices (the presence of the unity operator in the set of spin operators implied that we have algebra $u(2)$ instead of $su(2)$). From now on, we will assume summation over repeating Greek indices $\mu$, $\nu$ to run from $0,\ldots,3$.
In (\ref{2x2}) vectors $b_k$ are found to be 
\bea
\label{vectors}
b_1^0=\frac{(v_{00}+v_{11})}{2},\;b^3_1=\frac{(v_{00}-v_{11})}{2},\; b_1^+=b_1^-=v_{01},\hspace{0.4cm}\\
b_k^0=\frac{(v^2_{0k}+v^2_{1k})}{2},\;b^3_k=\frac{(v^2_{0k}-v^2_{1k})}{2},\; b_k^+=b_k^-=v_{0k}v_{1k}.\nn
\ena

One can see that Eq.~(\ref{vectors}) is surprisingly similar to one of the EKZ
equations, Eq.~(\ref{MKZ-2}), but instead of the full set of $n$ quantum spin generators,
we have only one spin generator $S_1^\mu$ associated with energy $\omega$
as a spectral parameter. Other $n-1$ spins, associated with spectral 
parameters $a_k, k=2,3,\cdots n$, are classical vectors, $b_k^\mu$. They can be treated quasi-classically as large quantum spin limits. The $b_1^\mu S_1^\mu$ term corresponds to $\lambda S^3$ modification in EKZ equations (\ref{MKZ-2}).

Now it is natural to ask whether there is a sufficient number of integrals of motion for the Hamiltonian in the Schr\"{o}dinger equation (\ref{2x2}), namely
\bea
\label{H2}
H_1=b_1^\mu S^\mu_1+\sum_{k=2}^n\frac{S_1^\mu b_k^\mu }{\omega-a_k},
\ena
to be integrable? One can expect that the set of operators
\bea
\label{Hk}
H_k=\sum_{k' > k}^n\frac{b^\mu_k b^\mu_{k'}}{a_k-a_{k'}}+\frac{b_k^\mu S_1^\mu}{a_k-\omega},\quad k=2,\cdots n
\ena
are candidates of integrals of motion, since they are large spin quasiclassical limits of the operators $H_j$ in the set of EKZ equations (\ref{MKZ-H}-\ref{MKZ}). Indeed, it appears that their commutators are equal to
 \bea
 \label{commuting}
 [H_i,H_j]&=&\frac{1}{(a_i-\omega)(a_j-\omega)}b_i^\mu b_j^\nu [S_1^\mu,S_1^\nu]\nn\\ 
 &=&\sum_{a,b,c=1}^3 \frac{b_i^a b_j^b \epsilon_{abc} S_1^c}{(a_i-\omega)(a_j-\omega)},\quad  i,j=1,2, \cdots n.
 \ena
These operators will commute if all $n$ classical limits of spins, $b^\mu_k$, $k=1,2,\cdots n$, are parallel: 
 $b_k || b_{k'}$. A straightforward analisys of expressions in Eq.~(\ref{vectors})
 for $b_k^\mu$ shows that the parall property will be fulfilled if
 \bea
 \label{parallelism}
 v_{ij}= \gamma_i \gamma_j,\quad i, j=0,1, \cdots n.
 \ena    
 Then we have a compete set of EKZ equations, where $\omega$ plays the role of the first 
 spectral parameter while others are $a_k$, $k=2,\cdots n$. Moreover,
EKZ  equations, Eq.~(\ref{MKZ}), with this set of Hamiltonians have a solution because
 zero curvature condition is fulfilled
 \bea
 \label{zero}
 \partial_{a_i}H_j-\partial_{a_j}H_i-[H_i,H_j]=0, \; i,j=0,2,\cdots n,
 \ena
 where $a_o=\omega$.
 Hence, the solution of the ADO model, with "parallel" conditions Eq.~(\ref{parallelism})
 on parameters, is defined by the solution of the EKZ equations with one quantum
 and $n-1$ classical spins.
  Wave function of ADO model, $\Phi(\omega, a_2,\cdots a_n)$, thus can be treated as correlation function $G(\omega,{a_i})$
 in EKZ equation.
 
  Interestingly, parallelism of $b^\mu_k$ vectors indicates that the $2\times 2$ matrix $b^\mu S^\mu_1=b_k^0 {\bf 1} + {\bf b}_k {\bf S}_1$ has rank one and the interaction term in Eqs.~(\ref{H2}-\ref{Hk}) coincides with the momentum map in Refs.~\cite{Adams-1990,Harnad-1994}. The latter  gives a dual characterization of integrable Hamiltonian systems as isospectral flow in the two loop algebras.

% \subsection{KZ integrability of the Demkov-Osherov model}
 \subsection{Solution of KZ equations}
 The solutions of KZ and EKZ equations based on OSABA were formulated in Refs.~\cite{Babujian-1994,Babujian-1996,Babujian-Kitaev}.
  The situation in the ADO model is simpler. 
  Since all vectors, $b_i$ are parallel it follows that all matrices
 $b^\mu_k S^\mu_1 $ are commuting. Therefore, one can work with them
 as with ordinary commuting numbers. 
 
 Due to this commutativity property, the inspection yields that
  \bea
  \label{sol-1}
  \partial_\omega (\omega-a_i)^{b^\mu S^\mu_1}=\frac{{b^\nu S^\nu_1}}{\omega-a_i}(\omega-a_i)^{b^\mu S^\mu_1}.
  \ena 
  Therefore, one can derive the general solution of EKZ equations based on the Hamiltonian $H_1$ given in Eq.~(\ref{H2}) and integrals of motion $H_i, i=2,\cdots n$ given by Eq.~(\ref{Hk}). Namely, the solution of 
   \bea
   \label{sol-2}
   \partial_{\omega}\Phi(\omega,\{a_i\})=-i H_1 \Phi(\omega,\{a_i\}),\nn\\
    \partial_{a_i}\Phi(\omega,\{a_i\})=-i H_i \Phi(\omega,\{a_i\}),
   \ena
we construct in several steps.  As the firs step, one can look for $\Phi(\omega,\{a_i\})$ in the form
	\bea
	\label{sol-3}
	\Phi(\omega,\{a_i\})&=&\prod_{j>i} (a_i-a_j)^{-i b_i^\mu b_j^\mu} \Phi^0(\omega,\{a_i\}).
	\ena
	Here vectors $b_k^\mu$ are defined by expressions provided in Eqs.~(\ref{vectors}) and (\ref{parallelism}). For the new "wavefunction" $\Phi_0$ we obtain a new equation that is similar to Eq.~(\ref{sol-2}), wherein the RHS of the second equation in the operator $ H_i $ one retains only the second term.
	As the second step,  in the same second equation one can separate the $\mu=0$ component of vectors $b_k^\mu$, $b_1^\mu$ from the rest in terms that contain them. This gives
	\bea
	\label{sol-4}
	\partial_{\omega}\Phi^0(\omega,\{a_i\})=\qquad\qquad\qquad\qquad\qquad\qquad\qquad\qquad\\
	-i \Biggl( {b_1^0 + {\bf b}_1{\bf S} + \sum_{k=2}^{n}}\frac{b_k^0}{\omega-a_k}
	+ \sum_{k=2}^{n}\frac{{\bf b}_k{\bf S}}{\omega- a_k}\Biggr) \Phi^0(\omega,\{a_i\}),\nn\\
	%    \label{sol-5}
	\partial_{a_i}\Phi^0(\omega,\{a_i\}) =-i\Big( \frac{b_i^0}{a_i-\omega} + \frac{{\bf b}_i{\bf S}}{a_i- \omega}\Big) \Phi^0(\omega,\{a_i\}).\nn
	\ena
	Here we define the ${\bf S} =(\sigma^1,\sigma^2, \sigma^3)$, where $\sigma^j$, $j=1,2,3$, are Pauli matrices.
	For $\Phi^0$ one will then obtain
		\bea
  \Phi_0(\omega,\{a_i\})= e^{-i \omega b_1^0}\prod_{k=2}^{n}(\omega -a_k)^{-i b_k^0}\Phi_s(\omega,\{a_i\}),\nn\\
	\ena
	where $\Phi_s(\omega,\{a_i\})$ is still unknown.
	After this one arrives at the equations for the function $\Phi_s(\omega,a_i)$ of the form
	\bea
	\label{sol-6}
	\partial_{\omega}\Phi_s(\omega,\{a_i\})&=&-i \left({\bf b}_1{\bf S}
	+  \sum_{k=2}^{n}\frac{{\bf b}_k{\bf S}}{\omega- a_k}\right) \Phi_s(\omega,\{a_i\})\nn\\
	% \label{sol-7}
	\partial_{a_i}\Phi_s &=&-i   \frac{{\bf b}_i {\bf S}}{a_i- \omega} \Phi_s(\omega,\{a_i\}).
	\ena
	Earlier we have observed that  commutativity of the Hamiltonians $H_i$ for different $i$ leads to the condition of all vectors ${\bf b}_i$ being parallel to each other, i.e.
	\bea
	\label{cond -1}
	{\bf b}_i= \beta_i {\bf n},
	\ena
	where ${\bf n}$ is an arbitrary unit vector and $\beta_i$ are the norms of  vectors ${\bf b}_i$. Below all these norms will be linked to the parameters of the ADO Hamiltonian. 
   
   Let us now rewrite Eqs.~(\ref{sol-4}) using the newly introduced notations:
	\bea
	\label{sol-6}
	\partial_{\omega}\Phi_s(\omega,\{a_i\})&=&-i \Big(\beta_1 {\bf n}{\bf S}  
	+ \sum_{k=2}^{n}\frac{{\beta_k \bf n}{\bf S}}{\omega- a_k}\Big) \Phi_s(\omega,\{a_i\}),\nn\\
	%    \label{sol-7}
	\partial_{a_i}\Phi_s(\omega,\{a_i\})& =&-i\Big(\frac{\beta_i {\bf n}
		{\bf S}}{a_i-\omega}\Big)\Phi_s(\omega,\{a_i\}).
	\ena     
	The eigenvectors and eigenvalues of ${\bf n}{\bf S}$ here are defined as   
	\bea
	\label{sol-7}  
	{\bf n}{\bf S} \xi_{m} &=& m\xi_{m},
	\ena
	where $\xi_{m}$ represents a two component spinor eigenvector with eigenvalue $m=\pm 1$. One can directly compute $\xi_{m}$  by solving eigenvalue equations directly in components. Using this fact we are now able to finally write the solution $\Phi_s^{m}$ in the form
	\bea
	\Phi_s^{m}=\exp(-i\omega\beta_1 m)\prod_{k=2}^{n}(\omega -a_k)^{-i\beta_km }\xi_{m}.
	\ena
	As the last step, we collect all parts of the solution and putting them together obtain
	\bea
	\label{sol-71}
	\Phi^{m}&=&\exp[-i\omega(b_1^0 +\beta_1m)] 
	\prod_{j>i} (a_i-a_j)^{-i (b_i^0 b_j^0 +\beta_i \beta_j)} \nn\\
	&\times&\prod_{k=2}^{n}(\omega -a_k)^{-i(b_k^0+\beta_km )}\xi_{m}.
	\ena
	As our equations are linear, the general solution will be a linear combination of the two
	different solutions corresponding to two eigenvalues with $m= \pm 1 $.

	One can express final answer in terms of parameters $\{\gamma_i\}$, $i=1,\ldots,n$ by use of the expressions in
	Eqs.~(\ref{vectors}) and (\ref{parallelism}) ensuring parallelism of $b$ vectors. In terms of parameters $\gamma$, the  $b$ vectors have the following form
	\bea
	\label{b}
	b_1^0&=&\frac{\gamma_0^2+\gamma_1^2}{2},\;\;\;\; b_1^3=\frac{\gamma_0^2-\gamma_1^2}{2},\;\;\;\;\; b_1^{\pm}=\gamma_0\gamma_1,\\
	b_k^0&=&\frac{\gamma_0^2+\gamma_1^2}{2}\gamma_k^2,\; b_k^3=\frac{\gamma_0^2-\gamma_1^2}{2}\gamma_k^2,\;b_k^{\pm}=\gamma_0\gamma_1\gamma_k^2,\nn
	\ena
from where it follows that
\bea
\label{sol-8}
\beta_1&=&b_1^0=\frac{\gamma_0^2+\gamma_1^2}{2},\; \\
\beta_k&=&b_k^0=\frac{\gamma_0^2+\gamma_1^2}{2}\gamma_k^2,\;\;\;\; k=2,3, \cdots N.\nn
\ena

Finally, by use of Eqs.~(\ref{b}) and (\ref{sol-8}), one can simplify the expression in Eq.~(\ref{sol-71}). Taking into account the factor $e^{i\omega^2/2}$
[we remind the reader about the transformation we introduced in the text below Eq.~(\ref{set})],the exact solution of ADO model acquires the form
\bea
\label{final}
\Phi(\omega, \{a_i\})&=& e^{i\omega^2/2}\prod_{j>i} (a_i-a_j)^{- 2 i \beta_i \beta_j}\\
&\times&\prod_{j=2}^n (\omega-a_j)^{-i \beta_j (1+m)} e^{-i \omega\beta_1(1+m)}\xi_m, \nn\\
\text{with}\quad m=\pm1 .\nn
\ena
This is one of the key results of the present work. Fourier transform of Eq.~(\ref{final}) for $m=\pm1$ defines the 
real-time evolution of the wave function
\bea
\label{function}
\varPhi(t,a_2,\cdots a_n)=\int_{-\infty}^\infty d\omega \Phi(\omega,a_2,\cdots a_n)e^{i \omega t}.
\ena
Appropriately normalized wave function gives the system's evolution
starting from the initial state at $t=-\infty$. Namely
\bea
\label{evolution}
\varPhi(t,a_2,\cdots a_n)=G(t,a_2,\cdots a_n)\varPhi(-\infty,a_2,\cdots a_n),
\ena  
where $G(t,a_2,\cdots a_n)=T\exp\{i \int_{-\infty}^t dt' H_{ADO} \}$
is the time evolution operator where $T$ stands for time ordering. It defines
the amplitudes of transmission probabilities between diabatic states at
$t=-\infty$ and  $t=\infty$. 
We can see from here that the state with $m=-1$ has trivial $\omega$ dependence, $\Phi(\omega)\sim e^{i\omega^2/2}$, which translates into a similar oscillatory time-dependence of the wave-function $\Phi(t)\sim e^{-it^2/2}$, that 
gives zero transmission probability between diabatic states. 

The analytical form of Fourier transformation of the wave function for arbitrary $n$ is cumbersome. However, for the simple case of $n=2$, it is straightforward, yielding the extract transmission probability $P$ of the Landau-Zener problem
\bea
\label{P}
P=e^{-2\pi (\gamma_0^2+\gamma_1^2)\gamma_2^2}.
\ena 
This is in full accordance with Landau-Zenner result of two band crossing\cite{Malla}, for parallel levels.

	\section{Conclusions}
	We have presented a set of quantum dynamical systems, which are linked to modified Knizhnik-Zamolodchikov equations and can be solved exactly upon employing this connection. Those are the BCS paring model (based on Richardson's exact solution), multi-level Landau-Zenner tunneling models, and their realizations as a generalization of Demkov-Osherov and bow-tie models. In these systems, interaction couplings can be considered to be time-dependent linearly, leading to the extension of KZ equations. This link is surprising and based on the fact that all these models contain integrals of motion of Gaudin magnets. This inherent property of integrability leads to EKZ equations. Using the link and integrable properties, we solve ADO and GBT models exactly. We believe this link of dynamical systems to EKZ equations is not limited and can be extended to other systems with time-dependent Hamiltonians. 
	
	Moreover, it is known that KZ equations are written for correlation functions of WZWN models. Here we have revisited the question posed in Ref.~\cite{Sedrakyan-2010}, namely what is the CFT for which EKZ defines correlation functions? Using a different method, we have shown that the corresponding model is the WZWN model with a boundary term discussed first in Ref. \cite{Sedrakyan-2010}.

\section*{Acknowledgement}
The authors are grateful to A. G. Sedrakyan and R. H. Poghossian for many useful discussions and to N. A. Sinitsyn for valuable comments.
The research was supported by startup funds from the University of Massachusetts, Amherst (T.A.S.), and H.M.B. was supported by the Armenian State Committee of Science in the framework of the research projects 191T-008, 20RF-142, and 21AG-1C024.

\appendix

\section{Appendix}
%\subsection{Current algebra relations}
\subsection{First term with derivative in the null-vector condition (\ref{null vector})}
From the operator algebra Eq.~(\ref{PF}), where  the current $J^a(u)$ is acting on  primary field  at the point $w_j$,  we have 
\bea
\label{J0} 
&&J_0^a \phi_{s_i}(w_i)=\oint_{C_i} du J^a(u) \phi_{s_i}(w_i)\nn\\
&=& \oint_{C_i} \frac{S_i^a}{u-w_i} \phi_{s_i}(w_i)=S_i^a \phi_{s_i}(w_i).
\ena	
Here we have used the fact that since $J^a(u)$ is acting on the primary field at $w_i$, the contour $C_i$ is circling the position $w_i$,
after which Cauchy integration is applied.

The first term in Eq.~(\ref{null vector}) reads
\bea
\label{A1}
&&\langle e^{\alpha \oint_{\cal C} dw w J^3(w)}\partial_{w_1}\phi_{s_1}(w_1)\cdots
\phi_{s_N}(w_N)\rangle\\
&=& \langle \sum_k \frac{\alpha^k}{k!}\Big( \oint_{\cal C}dw w J^3(w)\Big)^k\partial_{w_1}\phi_{s_1}(w_1)\cdots
\phi_{s_N}(w_N)\rangle\nn
\ena
Using Eq.~\ref{PF}, it is easy to calculate the linear in the current term in these series. Since, according to CFT, singularities in this operator product expansion may appear only at the positions $w_i$ of the primary field, we can shrink boundary contour ${\cal C}$ into the sum of circles $C_i$ around those points: ${\cal C}=\bigsqcup_i C_i$ (see Fig.\ref{contours}). Then each term defined by the contour $C_i$ gives the action of the current on primary field $\phi_{s_i}(w_i)$. Using Eq.~(\ref{PF}) for the action of current on the primary field, we obtain
\bea
\label{A2}
&&\langle \sum_{i=1}^N \oint_{C_i} dw w J^3(w)\partial_{w_1}\phi_{s_1}(w_1)\cdots
\phi_{s_N}(w_N)\rangle\nn\\
&=&\sum_{i=1}^N  \oint_{C_i} dw w \partial_{w_1} \frac{S_i^3}{w-w_i}\langle \phi_{s_1}(w_1)\cdots
\phi_{s_N}(w_N)\rangle\nn \\
&=& \sum_{i=1}^N \partial_{w_1}\big[ w_i S_i^3 \langle \phi_{s_1}(w_1)\cdots
\phi_{s_N}(w_N)\rangle\big]
\ena
Higher-order terms of current $J^3(w)$ will produce
series, where $\sum_i \oint_{C_i}dw w J^3(w)$ is replaced by $\sum_i w_i S_i^3$ and, therefore, one will get
\bea
\label{A3}
&&\langle e^{\alpha \oint_{\cal C} dw w J^3(w)}\partial_{w_1}\phi_{s_1}(w_1)\cdots
\phi_{s_N}(w_N)\rangle \nn\\
&=& \partial_{w_1}\big[ e^{\alpha \sum_{i=1}^N w_i S_i^3}\langle\phi_{s_1}(w_1)\cdots
\phi_{s_N}(w_N)\rangle\big].
\ena  
It is important to notice that the anomalous contractions $J^3(w) J^3(w') \sim \mathsf{k}/(w-w')^2$ between different currents in higher-order terms in the expansion of the exponent does not contribute into Eq.~(\ref{A3}). This happens because after regularization of multiple integrals $( \oint_{\cal C}dw w J^3(w) )^p$ by introducing infinitesimal shifts of $p$ contours, which is equivalent to the normal-ordered form of operators, one will not have poles within the smallest contour and thus the Cauchy integral will produce zero. 

Eq.~(\ref{A3}) reproduces Eq.~(\ref{term-1}) of the main text.

\subsection{Second term in the null-vector condition (\ref{null vector})}
 After expanding the exponent and using expressions (\ref{JJ}) for the currents, the second term in (\ref{null vector}) reads
 \bea
\label{A4}
I&=&\langle \Phi({\cal C})\big(J_{-1}^a J_0^a\big) \phi_{s_1}(w_1)\cdots
\phi_{s_N}(w_N)\rangle\nn\\
&=& \sum_{k=1}^\infty \frac{\alpha^k}{k!}\langle \big(\oint_{\cal C} dw w J^3(w) \big)^k \oint_{C_1}du \frac{J^a(u)}{u-w_1} S_1^a\nn\\
&\times& \phi_{s_1}(w_1)\cdots
\phi_{s_N}(w_N)\rangle
\ena
Here we have used the relation Eq.~(\ref{J0}) for the action of $J^a_0$ on primary fields and expression Eq.~(\ref{JJ}) for $J_{-1}^a$.
Eq.~\ref{A4} shows that besides primary fields at $w_i$, there are also currents $J^a(u)$ with which $J^3(w)$ in the exponent will have contractions according to $SU(2)$ current algebra relations
\bea
\label{CA}
J^3(w)J^{\pm}(u)=\pm \frac{J^{\pm}(u)}{w-u}.
\ena

\begin{figure}[h]
	\centerline{\includegraphics[width=60mm,angle=0,clip]{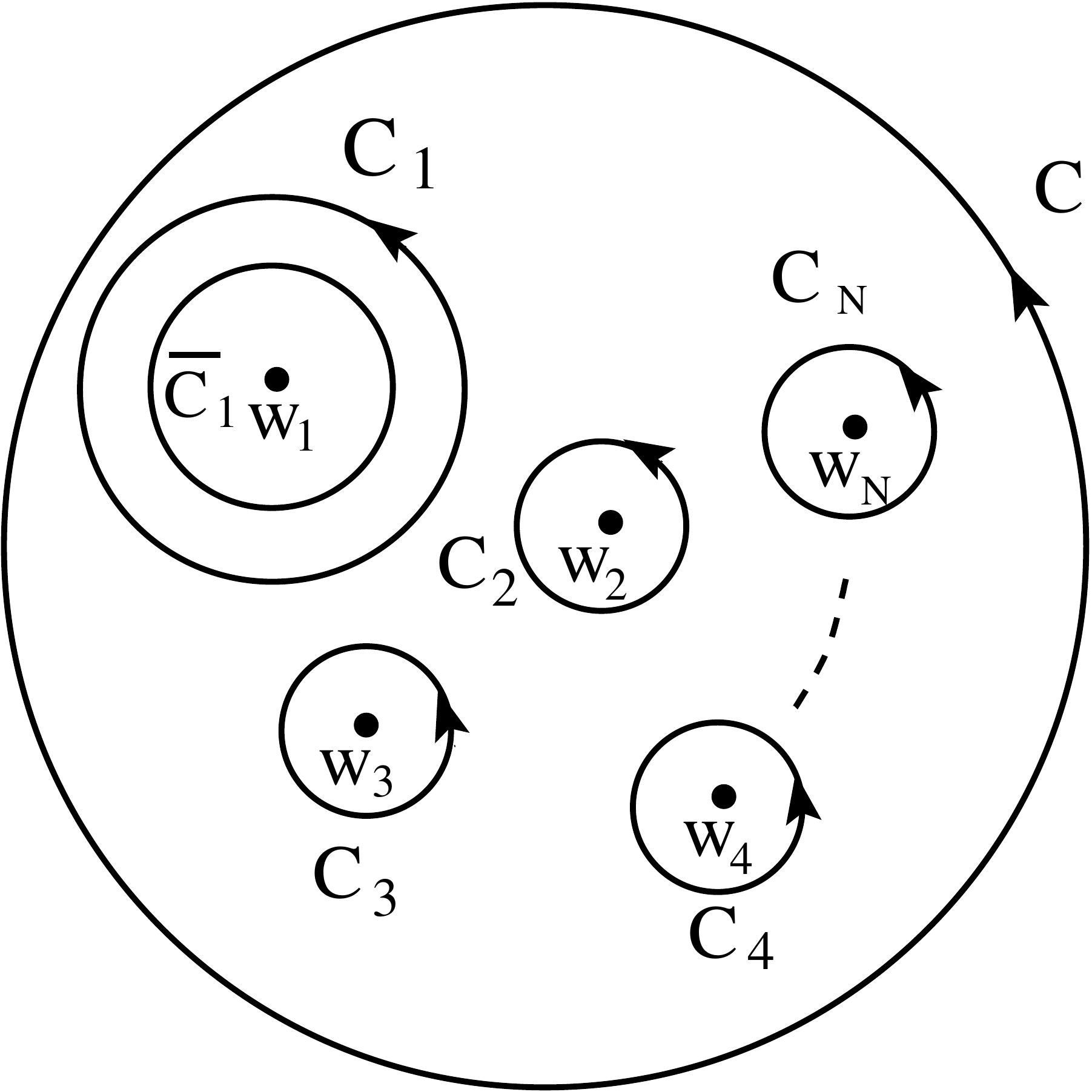}}
	\caption{Boundary contour {\cal C}  shrinks to the sum of contours $C_i,\;i=1,\cdots N$.  Here $\bar{C_1}$ and $C_i, \; i=2, \cdots N$ are the contours in the integral expression of the operators $J_{-1}^a$, which they act on the primary fields, $\phi_{s_i}(w_i)$, at points $w_i,\; i=1,\cdots N$. We have that $C_1 \supset \bar{C_1}$.} 
	\label{contours-u-2}
\end{figure}

 Therefore contour ${\cal C}$ will include singularities not only at $w_i$, $i=1 \cdots N$, but also in $u$ (see Fig.\ref{contours-u-2}).
 Hence boundary contour again will shrink into the sum ${\cal C}=\bigsqcup_{j=1}^N C_j$. 
 As in Eq.~(\ref{A3}), the contribution of the contours $C_j,\; 2=1,\cdots N$ is equivalent to the replacement of $\sum_{i=2}^N \oint_{C_i}dw w J^3(w)$  by $\sum_{i=2}^N w_i S_i^3$. As a result, from Eq.\ref{A4}, one obtains
 \bea
 \label{A5}
 I&=&  \langle \sum_{k=1}^\infty \frac{\alpha^k}{k!} \Big(\oint_{ C_1} dw w J^3(w) \Big)^k \oint_{\bar{C}_1}du \frac{J^a(u)}{u-w_1} S_1^a\nn\\
 &\times&  e^{\alpha \sum_{i=2}^N w_i S^3_i}\phi_{s_1}(w_1)\cdots \phi_{s_N}(w_N)\rangle
 \ena
The situation with $C_1$ is different. According to the CFT rules, the order of currents $J^3(w)$ and $J^a(u)$ defines the size of the contours in their integrals, namely, $C_1 \supset \bar{C_1}$, see Fig.\ref{contours-u-2}.

Now, let us first analyze the linear term in the series of exponential in
Eq.~(\ref{A5}). After using the current algebra relation Eq.`(\ref{A4}) and Cauchy integration over $w$, one obtains
 \bea
 \label{A6}
 &&\oint_{C_1}dw w \oint_{\bar{C_1}}du J^3(w)\frac{J^{\pm}(u)}{u-w_1}\nn\\
&=&\pm \oint_{C_1}dw w \oint_{\bar{C_1}}du \frac{J^{\pm}(u)}{(u-w_1)(w-u)}\nn\\
&=&\pm  \oint_{\bar{C_1}}du u \frac{J^{\pm}(u)}{(u-w_1)}.
 \ena  
 Hence, for the exponent in Eq.~(\ref{A4}), one will have
\bea
\label{A7}
e^{\alpha \oint_{C_1} dw w J^3(w)}\oint_{\bar{C_1}} du\frac{J^{\pm}(u)}{(u-w_1)}=
\oint_{\bar{C_1}}du \frac{e^{\pm u}J^{\pm}(u)}{(u-w_1)}
\ena
and the entire second term Eq.~(\ref{A4}) becomes
\bea
\label{A8}
I&=&\langle \oint_{\bar{C_1}}du\frac{e^{\alpha u}J^+(u)S^{-}_1+e^{-\alpha u}J^-(u)S^{+}_1}{u-w_1}\nn\\
&\times&  e^{\alpha \sum_{i=2}^N w_i S^3_i}\phi_{s_1}(w_1)\cdots \phi_{s_N}(w_N)\rangle
\ena
%%%%%%%%%%%%%%
%%%%%%%%%%%%%%
Using the extension of the relation Eq.~(\ref{PF}) for primary fields
to their product 
\bea
\label{PF-1}
J^a(u) \phi_{s_1}(w_1)\cdots \phi_{s_1}(w_N)&&\nn\\ 
= \sum_{i=1}^N\frac{S_i^a}{u-w_i} \phi_{s_i}(w_i)\cdots \phi_{s_1}(w_N).\
\ena 
and after Cauchy integration over $u$, one obtains
\bea
\label{A9}
I&=& \Big[\sum_{i=2}^N \frac{e^{\alpha w_1} S^+_i S^-_1+e^{-\alpha w_1} S^-_i S^+_1}{w_1-w_i} \nn\\
&+& \oint_{C_1}du  \frac{e^{\alpha u} S^+_1 S^-_1+e^{-\alpha u} S^-_1 S^+_1}{(u-w_1)^2}  \Big]\nn \\
&\times& e^{\alpha \sum_{i=2}^N w_i S^3_i}\langle \phi_{s_1}(w_1)\cdots \phi_{s_N}(w_N)\rangle
\ena
Now, after using the identity
\bea
\label{A10}
S_1^{\pm} e^{\mp \alpha w_1}=S_1^{\pm} e^{ \alpha w_1 S_1^3}
\ena
and performing Cauchy integration, one arrives at
\bea
\label{A11}
I&=&\Big[\sum_{i=2}^N \frac{S_i^a S_1^a}{w_1-w_i}+\alpha S_1^aS_1^a S_1^3\Big]\nn\\
&\times& e^{\alpha \sum_{i=2}^N w_i S^3_i}\langle \phi_{s_1}(w_1)\cdots \phi_{s_N}(w_N).
\ena
After defining $c_1=\sum_{a=1}^3 S_1^aS_1^a$, we see that Eq.~(\ref{A11}) coincides
with the relation Eq.~(\ref{term-2}) of the main text.
%Combining now two equations (\ref{A3}) and (\ref{A11})

%\newpage


\end{document}